\newif\ifarxiv
\arxivtrue   

\documentclass[11pt,a4paper,fleqn]{article}

\usepackage[
  a4paper,
  top=35.25 mm,
  bottom=25.4mm,
  left=25.4mm,
  right=25.4mm,
  headheight=14.06mm,
  headsep=20pt,
  footskip=12.5mm
]{geometry}
\usepackage{fontspec}

\ifarxiv
\else
\fi

\newfontfamily\dejavusans[
  Path=fonts/dejavusans/,
  Extension=.ttf,
  UprightFont=DejaVuSans,
  BoldFont=DejaVuSans-Bold,
  ItalicFont=DejaVuSans-Oblique,
  BoldItalicFont=DejaVuSans-BoldOblique
]{DejaVu Sans}

\renewcommand{\normalsize}{%
  \fontsize{11pt}{12.65pt}\selectfont
}
\normalsize

\usepackage{xcolor}
\definecolor{headergray}{gray}{0.3}

\usepackage{graphicx}   
\usepackage{amsmath}    
\usepackage{changepage} 
\usepackage{ragged2e}   
\usepackage{enumitem}
\usepackage{amssymb}
\usepackage[para,online]{threeparttable}

\ifarxiv
\else
  \usepackage{fancyhdr}
\fi

\usepackage{titlesec}
\titleformat{\section}
  {\bfseries\fontsize{12pt}{14pt}\selectfont} 
  {\thesection.}                               
  {0.5em}                                       
  {}                                            

\titlespacing*{\section}
  {0pt}{18pt}{12pt}  

\titleformat{\subsection}
{\bfseries\fontsize{12pt}{14pt}\selectfont}
{\thesubsection.}
{0.5em}
{}

\titlespacing*{\subsection}
{0pt}{12pt}{6pt}
\usepackage{caption}

\DeclareCaptionFont{tnr}{\fontsize{11pt}{12.65pt}\selectfont}

\newcommand{\figwidth}{0.75\linewidth}
\renewcommand{\arraystretch}{1.15}  

\usepackage[
  backend=biber,
  style=authoryear,
  sorting=nyt,
  maxcitenames=1,
  mincitenames=1,
  maxbibnames=3,
  giveninits=true,
  date=year,
  doi=false,
  url=false,
  isbn=false,
  eprint=false,
  uniquename=false,
  uniquelist=false
]{biblatex}

\AtEveryBibitem{%
  \clearfield{urldate}%
  \clearfield{url}%
  \clearfield{note}%
  \clearfield{addendum}%
}

\AtBeginBibliography{%
  \setlength{\bibhang}{0pt}%
  \setlength{\itemindent}{0pt}%
}

\ifarxiv
    \usepackage[frozencache,cachedir=minted-cache]{minted}
\else
    \usepackage[cachedir=minted-cache]{minted}
\fi
\setminted{
    autogobble=true,
    breaklines=true,
    fontsize=\tiny
}

\makeatletter
\renewcommand{\maketitle}{%
  \begin{center}
    \vspace*{24pt}
    {\bfseries\fontsize{14pt}{16pt}\selectfont \@title \par}
    \vspace{30pt}
    {\fontsize{11pt}{12.65pt}\selectfont \@author \par}
    \vspace{17.3pt} 
  \end{center}

  {
      \setlength{\topsep}{0pt} 
      \begin{adjustwidth}{1cm}{1cm} 
        {\fontsize{11pt}{12.65pt}\selectfont 
          \justifying           
          \noindent
\noindent\textbf{Abstract:}
Agentic workflows driven by large language models (LLMs) are increasingly applied to Building Information Modelling (BIM), enabling natural-language retrieval, modification and generation of IFC models. Recent work has begun adopting the emerging Model Context Protocol (MCP) as a uniform tool-calling interface for LLMs, simplifying the agent side of BIM interaction. While MCP standardises how LLMs invoke tools, current BIM-side implementations are still authoring tool-specific and ad hoc, limiting reuse, evaluation, and workflow portability across environments. This paper addresses this gap by introducing a modular reference architecture for MCP servers that enables API-agnostic, isolated and reproducible agentic BIM interactions by design. Based on a systematic analysis of recurring capabilities in recent literature, we derive a core set of requirements. These inform a microservice architecture centred on an explicit adapter contract that decouples the MCP interface from specific BIM APIs at the architectural level. We demonstrate the practicality of both implementation and use via an IFCOpenShell-based prototype. An evaluation across common modification and generation tasks demonstrates feasibility and suggests that the architecture provides a reusable foundation for systematic research. 

\vspace{1em}%
        }
      \end{adjustwidth}
  }
}
\makeatother

\title{\MakeUppercase{A Modular Reference Architecture for MCP-Servers Enabling Agentic BIM Interaction}} 

\author{
Heimig-Elschner, Tobias$^{1,3}$, Du, Changyu$^{2,4}$, Scheuvens, Anna$^{1}$,  Borrmann, André$^{2,4}$,  Beetz, Jakob$^{1}$\\ 
$^{1}$Chair of Design Computation, RWTH Aachen University, Germany\\
$^{2}$Chair of Computing in Civil and Building Engineering, Technical University of Munich, Germany\\
$^{3}$Federal Institute for Research on Building, Urban Affairs and Spatial Development (BBSR), Germany\\
$^{4}$TUM Georg Nemetschek Institute, Germany\\
E-mail: Tobias@heimig.de
}

\date{}  

\begin{document}

\maketitle
\noindent

\section{Introduction} 
\label{sec:introduction}

Recent advances in AI — particularly large language models (LLMs) and agent-based systems — are transforming knowledge-intensive domains by enabling automation, decision support, and multimodal reasoning \parencite{kar_2023_UnravellingImpactGenerative, ng_2021_SystematicLiteratureReview}. Within the Architecture, Engineering, and Construction (AEC) industry, this trend coincides with the continued shift towards model-centric workflows built on Building Information Modelling (BIM). The ISO-standardized Industry Foundation Classes (IFC) provide an open, semantically rich data model that enables reliable and interoperable exchange across heterogeneous BIM software ecosystems \parencite{borrmann_2021_BuildingInformationModeling}.

In this context, the interaction of LLM-based agents with BIM models has emerged as a promising research direction. Recent studies show that agents can retrieve information from IFC models \parencite{fernandes_2024_GPTPoweredAssistantRealTime, li_2025_InteractiveSystem3D, Hellin2025_EC3}, support modelling tasks \parencite{du_2025_Text2BIMGeneratingBuilding, jang_2024_AutomatedDetailingExterior}, and perform design reasoning when coupled with BIM-specific APIs \parencite{du_2024_CopilotBIMAuthoring}. These approaches consistently rely on programmatic tool calls and integrations with BIM authoring environments, forming an emerging foundation for structured and reproducible agentic workflows.

Despite this progress, existing implementations remain fragmented and repeatedly re-implement the same core capabilities, limiting generalisability and reuse. This paper addresses this gap by introducing a modular reference architecture that uses the Model Context Protocol (MCP) to unify agent–tool interaction and provide a reusable foundation for retrieval, modification, and generation workflows across different BIM APIs.
\section{State of the Art} 
\label{sec:stateoftheart}
\subsection{Agentic BIM \& LLM Workflows} 

Emerging research has integrated LLMs into BIM workflow as single- or multi-agent systems, enabling natural-language retrieval, generation, and model modifications. For information retrieval, \cite{Hellin2025_EC3} proposed an IFC-based multi-agent query system that achieves high accuracy by iteratively calling tools to interact with models. \cite{Avgoren:2025:IFCLLMGraph} mapped IFC to a semantic knowledge graph and used LLM-generated Cypher query to support Q\&A with 3D visual feedback. For the model generation and modification, \parencite{du_2025_Text2BIMGeneratingBuilding} translates user requirements into executable BIM software API calls via multiple LLM agents and uses a model checker for iterative refinement. Similarly, \parencite{wei2025text} proposed a system for text-to-code generation of modular building layouts, and \parencite{dong_2025_AIBIMCoordinator} proposed a multi-agent system to support various BIM design coordination tasks. Overall, existing studies follow a compact pattern: LLMs plan and orchestrate local API-based tools (e.g., commercial authoring software/open-source APIs), with lightweight iterative checking to improve correctness and control.

\subsection{The Model Context Protocol and its application in agentic BIM}

The MCP is a recent open standard for connecting LLM-based agents and external tools through a uniform JSON-RPC interface \parencite{anthropic_mcp_2024, mcp_spec_2025}. It replaces ad-hoc integrations with a consistent mechanism for tool description, discovery, invocation, and structured context provision, enabling reusable and vendor-agnostic agent–tool workflows \parencite{googlecloud_mcp_2025}. It cleanly separates the LLM host from the executing tool server, allowing backends to be swapped or extended without modifying the agent. Key technical features include uniform schemas, streaming support (STDIO/SSE), and explicit permission and isolation mechanisms \parencite{mcp_spec_2025}.

A small but growing ecosystem of MCP servers leverages this standard for BIM/IFC interaction. Existing implementations span open BIM environments such as Bonsai, WebIFC/Fragments, and IfcOpenShell, as well as emerging prototypes for proprietary tools including Autodesk Revit and Graphisoft Archicad. An overview of accessible implementations is provided in Table~\ref{tab:mcp_servers}. Most servers follow a monolithic design in which tool invocation, IFC processing, and model updates are executed within a single tightly integrated runtime, typically exposed via a Stdio-based MCP interface and connected to local BIM backends through Python bindings, TCP bridges, or IPC layers. Despite these architectural constraints, current implementations already support common agentic workflows such as element querying, model inspection, editing, and procedural scene construction.

\begin{table}[h]
\centering
\begin{threeparttable}
\caption{MCP server implementations for agentic BIM interaction}
\label{tab:mcp_servers}

\begin{tabular}{l l c c l}
\hline
\textbf{Name} & \textbf{BIM Authoring} & \textbf{Open-BIM} & \textbf{Monolithic} & \textbf{Protocol} \\
\hline
Bonsai-MCP & Bonsai & $\checkmark$ & $\checkmark$ & stdio \\
ifcMCP & IFCOpenShell & $\checkmark$ & $\checkmark$ & HTTP (streamable) \\
MCP4IFC\tnote{*} & Bonsai & $\checkmark$ & $\checkmark$ & stdio \\
Fragment MCP & Web-IFC & $\checkmark$ & $\checkmark$ & stdio \\
Tapir-MCP & Archicad & $\times$ & $\checkmark$ & stdio \\
revit-mcp & Revit & $\times$ & $\checkmark$ & stdio \\
xml.Revit.MCP & Revit & $\times$ & $\checkmark$ & stdio \\
Revit MCP (Beta)\tnote{**} & Revit & $\times$ & ? & Unknown \\
\hline
\end{tabular}

\begin{tablenotes}
\footnotesize
\item[*] \parencite{nithyanantham_2025_MCP4IFCIFCBasedBuilding}; \item[**] Only beta announced.
\end{tablenotes}

\end{threeparttable}
\end{table}

Among these efforts, MCP4IFC remains the sole framework reported in the scientific literature \parencite{nithyanantham_2025_MCP4IFCIFCBasedBuilding}. It introduces a layered tool organisation and demonstrates IFC querying, modification, and stepwise model generation using a combined Bonsai–IfcOpenShell backend. Together, these systems represent the first generation of MCP-based BIM servers and serve as the foundation for the modular, backend-agnostic reference architecture developed in this work.
\section{Methodology} 
\label{sec:methodology}

\subsection{Research questions and design}
Recent work on agentic BIM workflows demonstrates substantial progress in model manipulation, information retrieval, and validation. However, existing implementations remain fragmented, monolithic and tightly coupled to specific BIM authoring tools or APIs. This capability gap directly motivates the following research questions:

\textbf{RQ1:} Which core capabilities required for agentic BIM interaction can be identified from existing LLM-based workflows, and how can these be consolidated into a reusable reference architecture?\\
\textbf{RQ2:} Which architectural principles enable a modular, BIM-API-agnostic, and tool-isolated MCP-based system design, and how can these principles be realized in practice?\\
\textbf{RQ3:} To what extent can the proposed reference architecture reliably support typical agentic BIM workflows - specifically model modification and model generation?

The study follows a Design Science Research (DSR) approach, in line with \cite{hevner_2004_DesignScienceInformation}, which structures research around the iterative development and evaluation of artefacts. In this work, DSR is implemented by identifying recurring capability needs in recent agentic BIM workflows, designing a modular MCP-based reference architecture to address them, instantiating this architecture through a prototype built around an isolated IfcOpenShell execution backend, and evaluating it through scenario-based tests focusing on modification and generation tasks. Therefore, a simplified variant of \cite{du_2025_Text2BIMGeneratingBuilding}'s Text2BIM agent is re-implemented, and test cases are assessed to capture complementary aspects of agent behaviour, model validity, and model quality.

\subsection{Architectural Design Principles}
Grounded in the systematic analysis of existing agentic BIM workflows and emerging MCP-servers for BIM authoring tools and their recurring capability requirements, the proposed architecture follows a set of design principles that guided artefact development. These principles emphasize:  \textbf{(i)} Modular microservice decomposition of core capabilities to enable extensibility and substitution of components isolating the generalized MCP-server from the specific tooling; \textbf{(ii)} API-agnosticism to support heterogeneous BIM execution backends; \textbf{(iii)} Isolation of the BIM execution layer for safety and backend flexibility; \textbf{(iv)} Reusable, generic MCP tools at both low and high levels to minimize repeated implementation effort.


\section{Reference Architecture} 
\label{sec:ref-architecture}

The proposed reference architecture consolidates the recurring capabilities identified in prior work and operationalizes them through a set of decoupled microservices interconnected via HTTP interfaces and exposed through the MCP as a unified, standardized interface for agentic systems. The architecture is intentionally BIM-authoring-tool-agnostic, tool-isolated, and extensible, addressing the fragmentation gap summarized in Section~\ref{sec:stateoftheart}.

\subsection{Problem Identification: Common core capabilities and tools}
\label{subsec:core-capabilities}
As shown in Section~\ref{sec:stateoftheart}, among a set of relevant studies on agentic BIM interaction a common core set of tools and generalizable capabilities can be identified. Those capabilities can be condensed to the following seven:
\begin{enumerate}[label=(\roman*), align=left, itemsep=2pt]
    \item \textbf{Model File \& Storage Management (C1):} Handling import and export, versioning, persistence and access of BIM models (e.g., IFC, authoring tool formats, cloud storage)
     \item \textbf{BIM Execution (C2):}	Executing operations on BIM models - querying, creation, deletion, modification of elements (walls, volumes, layouts), updating geometry and semantics
    \item \textbf{Knowledge \& Context Provision (C3):} Providing domain knowledge and semantics to inform agent decisions - BIM workflows, API documentations, building code rules and user context
    \item \textbf{User-Model-Interaction (C4):}	Enabling user interaction with the current BIM model due to  visualization, upload and download
    \item \textbf{Information Retrieval \& Exploration (C5):} Extracting and navigating information within a BIM model - semantic queries, spatial relationships, retrieving metadata, answering questions
    \item \textbf{Model Validation \& Quality Assurance (C6):} Checking correctness, compliance, rule-based verification of BIM models, detecting errors/hallucinations, ensuring model quality
    \item \textbf{Multi-Modal Inputs \& Outputs (C7):} Supporting other modalities beyond text: voice commands, sketches, images, sensors, AR/VR
\end{enumerate}

As summarised in Table~\ref{tab:capabilites}, these capabilities are consistently implemented across the analysed studies; however, they are almost always realised directly within the respective BIM authoring tool, resulting in a strong coupling between agentic logic and the underlying BIM API.\\

\begin{table}[h]
\centering
\begin{threeparttable}
\caption{Capability implementation in agentic BIM workflows}
\label{tab:capabilites}


\begin{tabular}{l l l l l l l l}

\hline
 & \textbf{C1} & \textbf{C2} & \textbf{C3} & \textbf{C4} & \textbf{C5} & \textbf{C6} & \textbf{C7} \\
\hline
\cite{du_2025_Text2BIMGeneratingBuilding}  & $\checkmark$ & $\checkmark$ & $\checkmark$ & $\checkmark$ & $\times$ & $\checkmark$ & $\times$ \\
\cite{du_2024_CopilotBIMAuthoring}  & $\checkmark$ & $\checkmark$ & $\checkmark$ & $\checkmark$ & $\checkmark$ & $\times$ & $\checkmark$ \\
\cite{deng_2025_BIMgentAutonomousBuilding} & $\checkmark$ & $\checkmark$ & $\checkmark$ & $\checkmark$ & $\checkmark$ & $\checkmark$ & $\checkmark$ \\
\cite{jang_2024_AutomatedDetailingExterior}  & $\checkmark$ & $\checkmark$ & $\times$ & $\checkmark$ & $\checkmark$ & $\times$ & $\checkmark$ \\
\cite{duggempudi_2025_TexttoLayoutGenerativeWorkflow}  & $\times$ &  $\checkmark$ & $\checkmark$ & $\times$ & $\times$ & $\checkmark$ & $\times$ \\
\cite{dong_2025_AIBIMCoordinator}  & $\checkmark$ & $\checkmark$ & $\checkmark$ & $\checkmark$ & $\checkmark$ & $\checkmark$ & $\checkmark$ \\
\cite{fernandes_2024_GPTPoweredAssistantRealTime}  & $\checkmark$ & $\checkmark$ & $\checkmark$ & $\checkmark$ & $\checkmark$ & $\times$ & $\checkmark$ \\
\cite{zheng_2023_DynamicPromptbasedVirtual} & $\checkmark$ & $\times$ & $\checkmark$ & $\checkmark$ & $\checkmark$ & $\times$ & $\checkmark$ \\
\cite{Hellin2025_EC3} & $\times$ & $\checkmark$ & $\checkmark$ & $\checkmark$ & $\checkmark$ & $\times$ & $\times$ \\
\cite{li_2025_InteractiveSystem3D} & $\times$ & $\times$ & $\checkmark$ & $\checkmark$ & $\checkmark$ & $\times$ & $\checkmark$ \\
\cite{liu_2025_BIMCoderComprehensiveLarge} & $\times$ & $\times$ & $\checkmark$ & $\checkmark$ & $\checkmark$ & $\checkmark$ & $\times$ \\
\cite{koh_2026_CosteffectiveMinimalinterventionBIM} & $\times$ & $\times$ & $\checkmark$ & $\checkmark$ & $\checkmark$ & $\checkmark$ & $\times$ \\
\hline
\end{tabular}

\begin{tablenotes}
\footnotesize
\end{tablenotes}

\end{threeparttable}
\end{table}

Beyond these capability categories, the analysed studies also reveal a recurring tool structure. Broadly, two classes of tools can be distinguished: (i) general-purpose tools (e.g., arithmetic or text utilities), and (ii) BIM-specific tools operating on the underlying model. Following the taxonomy introduced by \cite{nithyanantham_2025_MCP4IFCIFCBasedBuilding}, BIM-oriented tools can be organised around the fundamental operations \emph{create}, \emph{query}, and \emph{modify}. They can further be separated into \emph{low-level} functions, which execute API-specific code directly on the BIM model, and \emph{high-level} functions that encapsulate common modelling tasks. 
While \cite{nithyanantham_2025_MCP4IFCIFCBasedBuilding} already demonstrate abstraction and summarisation mechanisms — and similar tendencies can be observed in emerging BIM-tool-specific MCP-servers — all existing implementations and their tool abstractions remain tightly coupled to a single BIM authoring backend.
As a result, each system re-implements similar capabilities using backend-specific logic, even for tasks not inherently tied to BIM operations (e.g., file handling, user-interaction or information provisioning). This prevents such capabilities from being replaced, extended, or shared across backends and contributes to a fragmented ecosystem limiting extensibility, comparability, and systematic advancement of agentic BIM research—even when MCP is employed as a unifying interface.

\subsection{Solution Design: Architecture Overview}

\begin{figure}[h]
  \centering
  \includegraphics[width=\figwidth]{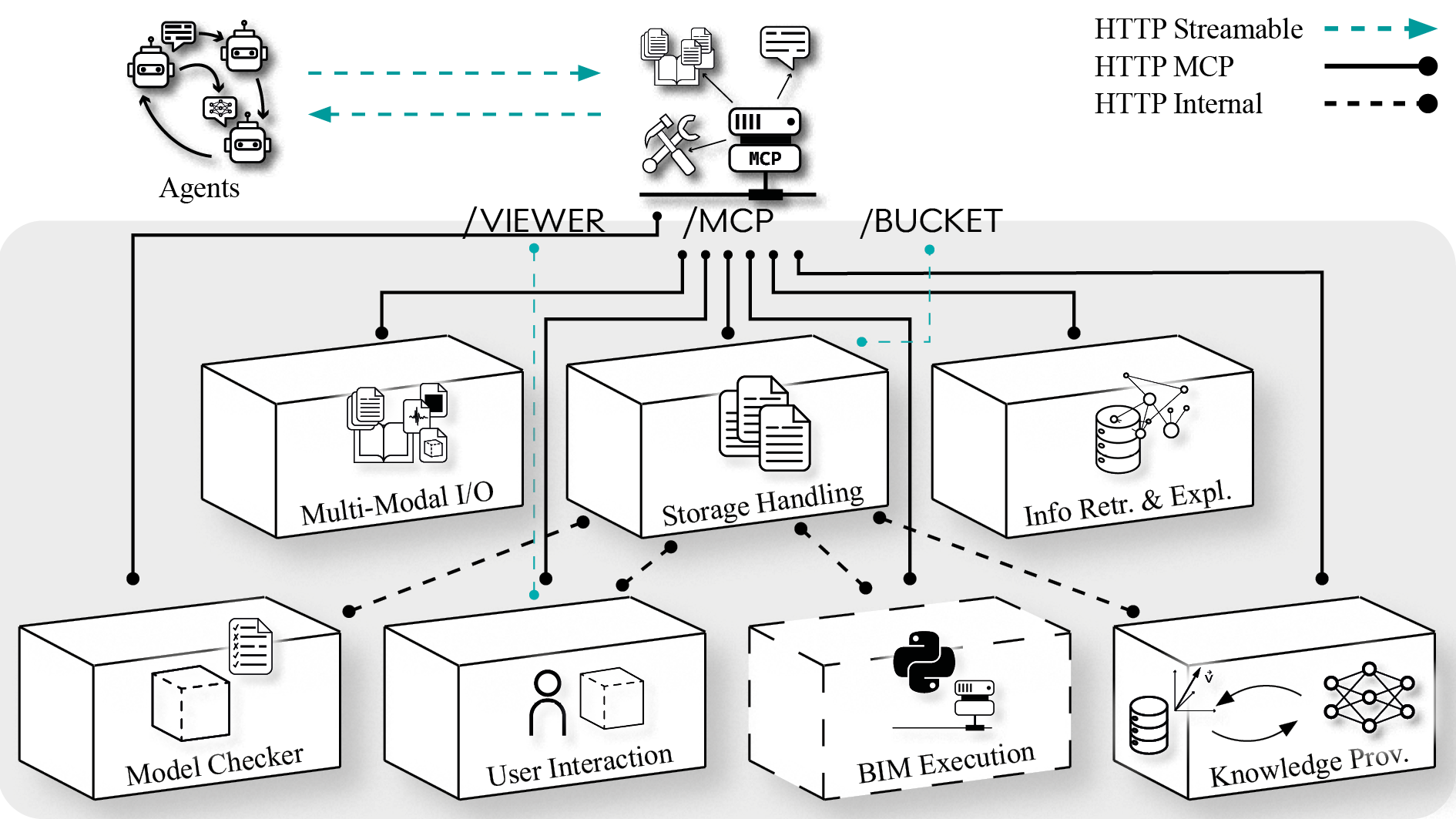}
  \caption{Conceptual visualization of the container microservice architecture.}
  \label{fig:ref_arch}
\end{figure}

To address the limitations identified in current MCP-based BIM implementations—namely the tight coupling between MCP servers and single BIM authoring tools—we propose a modular microservice reference architecture that abstracts core capabilities into isolated, exchangeable services. The architecture (Figure~\ref{fig:ref_arch}) positions the Model Context Protocol (MCP) as the unified agent-facing interface, while delegating all capability-specific logic to dedicated containerised services that communicate internally via HTTP endpoints. Overall, this design addresses key fragmentation issues and provides a reusable, architectural foundation for systematic evaluation and extensibility across heterogeneous BIM backends.

At its core, the system comprises three exposed entry points: (i) a standardized MCP endpoint (``/mcp'') implementing streamable HTTP for agent interaction, (ii) a viewer endpoint (``/viewer'') offering geometry visualisation, uploading, and interactive inspection for human–agent workflows, and (iii) a bucket endpoint (``/bucket'') responsible for storing, versioning, and exchanging IFC models and auxiliary artefacts \ref{fig:ref_arch}. All remaining components operate behind these entry points as independent microservices, each realising one of the previously identified capability categories. Furthermore, communication via (streamable) HTTP enables the MCP server to be deployed either locally or remotely, providing flexibility in scaling, distribution, and execution.

\subsubsection{BIM execution isolation}

A central architectural element is the isolation of the BIM Execution Service, which is responsible for all API-specific BIM operations and constitutes the primary source of coupling in existing MCP-based implementations. In the proposed framework, this execution layer is encapsulated in a standalone, sandboxed container that exposes only three HTTP endpoints (``/query``, ``/create``, ``/modify``), each defined by a formal JSON schema governing inputs and outputs. The MCP server never executes BIM-API code directly; instead, all model operations are delegated to the execution service through a strict adapter contract implemented as an abstract base class following a minimal capability pattern derived from common agentic BIM workflows. This contract specifies the interaction protocol for loading and saving models, performing element- and geometry-level operations, and computing differences. Thus, the heterogeneous BIM backends (e.g., IfcOpenShell, Revit, Archicad) can be integrated through backend-specific adapter implementations that satisfy a required core set of capabilities while preserving a uniform MCP tool surface.
The execution container evaluates backend-specific logic, isolated from the MCP server’s memory and filesystem. Model inputs and outputs are resolved through S3-compatible, versioned object storage, while the service itself remains stateless across requests. Each operation returns a structured \emph{Artifact} containing the updated file reference, a \emph{Manifest} with timestamps and tool metadata, an optional \emph{logicalResult} for query operations, and a detailed \emph{DiffRaw}/\emph{DiffSummary} capturing all IFC-level modifications. By externalising all state to a versioned storage layer, the framework enables reproducibility, fine-grained inspection of intermediate states, and multi-agent collaboration.

\subsubsection{Interaction and Execution Model}

The interaction loop between the agent, the MCP server, and the BIM Execution Service follows a ReAct-style pattern: the agent receives a task, contextual information, the active tool catalogue, and a reference to the current model version. Based on that, it reasons and issues a tool call through the standardized MCP endpoint. Each tool invocation is translated by the MCP server into a backend-agnostic operation expressed solely through the abstract interfaces defined by the \texttt{BaseAdapter} contract and is then forwarded to the isolated execution container. Within this container, the concrete adapter implementation converts the generalised instruction into backend-specific logic - such as API calls or executable code fragments - and performs the operation on the referenced model. 

The execution service processes the request by loading the referenced model, applying the backend-specific operation, computing an IFC-level diff, and writing the updated artefact back to the versioned storage system. It returns a structured \texttt{Artifact} containing an updated file reference, a \texttt{Manifest} with metadata, an optional \texttt{logicalResult} for queries, and detailed \texttt{DiffRaw} and \texttt{DiffSummary} representations. The MCP server converts this response into a compact \texttt{ChatArtifact} and injects it into the agent’s context, updating the model state accessible to subsequent reasoning steps. This interaction model ensures that the agent operates over a transparent sequence of model states, each accompanied by structured metadata and diffs. It supports reproducible multi-step reasoning, fine-grained inspection of intermediate modifications, and consistent behaviour across heterogeneous BIM backends, while keeping the MCP server free of any backend-specific execution logic.

\subsubsection{MCP Tool Hierarchy}

Following the taxonomy proposed by \cite{nithyanantham_2025_MCP4IFCIFCBasedBuilding}, the MCP server organises its tool catalogue analogically reflecting the three fundamental BIM operations (query, create, modify) and the modular composition of the reference architecture. At the foundation, a set of \emph{low-level BIM tools} exposes a direct, backend-agnostic interface to the standardized endpoints of the BIM Execution Service. These tools allow the execution of abstract methods defined by the \texttt{BaseAdapter} contract as well as backend-specific code fragments inside the execution container. Building on these primitives, the MCP server provides \emph{high-level BIM tools} that bundle recurring modelling tasks and interaction patterns into reusable abstractions. High-level tools do not manipulate IFC files or API objects directly; they emit backend-agnostic instructions that the execution container resolves using its specific adapter implementation. Beyond BIM-oriented tools, additional toolsets associated with other microservices in the architecture, including documentation and API retrieval (knowledge service), semantic and graph-based information access (semantic service), and tools for model upload, download, and inspection (viewer service) are exposed. All tools return results in the unified \emph{Artifact} and \emph{Manifest} format, supporting traceability, reproducibility, and coherent multi-step agent reasoning. While using \textit{low\_level} -tools the execution environment provides a controlled form of “semantic sugar” by injecting the concrete adapter instance as a pre-initialised global object (\texttt{adapter}), allowing access to helper functions, placement logic, and semantic utilities while maintaining backend isolation.

\section{Implementation} 

A prototype of the proposed modular architecture was implemented using a set of lightweight microservices communicated via HTTP and presigned S3 URLs. The implementation realises four of the seven capability categories --- \emph{BIM Execution}, \emph{File \& Storage Management}, \emph{Knowledge \& Context Provision}, and \emph{User Interaction} --- with a focus on BIM model generation and modification. Table~\ref{tab:containers_capabilities_short} summarises the correspondence between the Docker services and the capability categories from Section~\ref{sec:ref-architecture}. All source code is available on GitLab\footnote{\url{https://gitlab.com/phd5392441/mcp4bim/-/tree/882e3a1c8c34fbd5321ac4f4510e2063990b6276/}}. The agent-facing MCP server is implemented using \texttt{FastMCP} on top of FastAPI, exposing a streamable HTTP endpoint (``/mcp''). The server registers a structured tool catalogue comprising low\- and high\-level BIM tools. Tool definitions follow the architecture described in Section~\ref{sec:ref-architecture} and are implemented in the module \texttt{tools}. All tool outputs are encoded as compact \emph{ChatArtifacts}, derived from the Pydantic \emph{Artifact} and \emph{Manifest} schema. These carry file references, diffs and logical results while keeping context size minimal, enabling reproducible multi-step agent workflows.

\begin{table}[h]
\centering
\caption{Mapping of services to capability categories.}
\label{tab:containers_capabilities_short}
\begin{tabular*}{\linewidth}{@{}l p{0.8\textwidth}@{}}
\renewcommand{\arraystretch}
\hline
\textbf{Service} & \textbf{Implemented Capabilities} \\
\hline
\texttt{mcp-server} & Agent interface; tool provisioning; service orchestration. \\
\texttt{bim-exec} & BIM Execution (create/query/modify) \& diffing (backend-specific). \\
\texttt{minio} & File \& Storage Management; IFC versioning. \\
\texttt{weaviate \& ollama} & Knowledge \& Context Provision via vector-based retrieval. \\
\texttt{viewer-service} & User Model--Interaction: visualisation, upload/download. \\

\end{tabular*}
\end{table}
All BIM-API-specific logic is isolated in a dedicated \texttt{BIM\_EXEC} container implemented with Flask. The service exposes three JSON-schema-validated endpoints --- ``/create``, ``/modify`` and ``/query`` --- as defined in the request schemas. Each request triggers a sandboxed Python execution: input models are downloaded via presigned GET URLs, backend-specific code is executed, IFC-level diffs are computed, and updated models are uploaded to MinIO via presigned POST URLs.
Backend functionality is provided by an \texttt{IFCOSAdapter}, an implementation of the abstract \texttt{BaseAdapter} contract. The adapter defines loading, saving and diffing of IFC files and implements high-level IFC creation functions. Diffs are produced using the \texttt{entity\_diff} helper and returned as \texttt{DiffRaw}/\texttt{DiffSummary}, enabling transparent reasoning over model changes.

\section{Evaluation} 
\label{sec:evaluation}
This section assesses whether the proposed architecture can reliably support multi-step BIM generation workflows. The evaluation is explicitly scoped as a feasibility and non-regression study, demonstrating that the modular architecture can robustly realise existing agentic BIM workflows without loss of functionality. Accordingly, the focus lies not on comparative performance analysis, prompt engineering, or empirical validation of microservice-specific properties such as scalability or service exchangeability.

\subsection{Evaluation Methodology}
The architecture is evaluated using artefact-centred, scenario-based workflows focusing on generation tasks. A simplified variant of \cite{du_2025_Text2BIMGeneratingBuilding}'s Text2BIM multi-agent system is implemented as a single-agent ReAct-style setup \parencite{yao_2022_ReActSynergizingReasoning} using \texttt{gpt-5-mini} as the underlying language model. 
Unlike the original Text2BIM implementation, which involves four dedicated agents collaboratively generating Python scripts that invoke the BIM authoring software's APIs, the current implementation utilizes an adjusted system prompt, lightweight history management, and exclusively accesses all BIM-related functionality through the MCP server. Additionally, the model checking loop from the original implementation is also omitted here to maintain the scope of this study.

The evaluation uses six predefined test cases, combining two newly designed cases and four adapted from the original Text2BIM study \parencite{du_2025_Text2BIMGeneratingBuilding}. Each test case is executed under clean initial conditions and repeated five times to assess the stability and reproducibility of agent behaviour. Metrics cover three complementary aspects: \textit{Agent Metrics}, \textit{Model Validation Metrics} and \textit{Model Quality Metrics}. Model quality, understood as conformity with the described design requirements, is assessed via rule-based checks complemented by manual revues using a 5-level scale (Appendix). 

\subsection{Results}
\begin{figure}[htb]
  \centering

  \begin{minipage}[t]{0.3\textwidth}
    \centering
    \includegraphics[width=\linewidth]{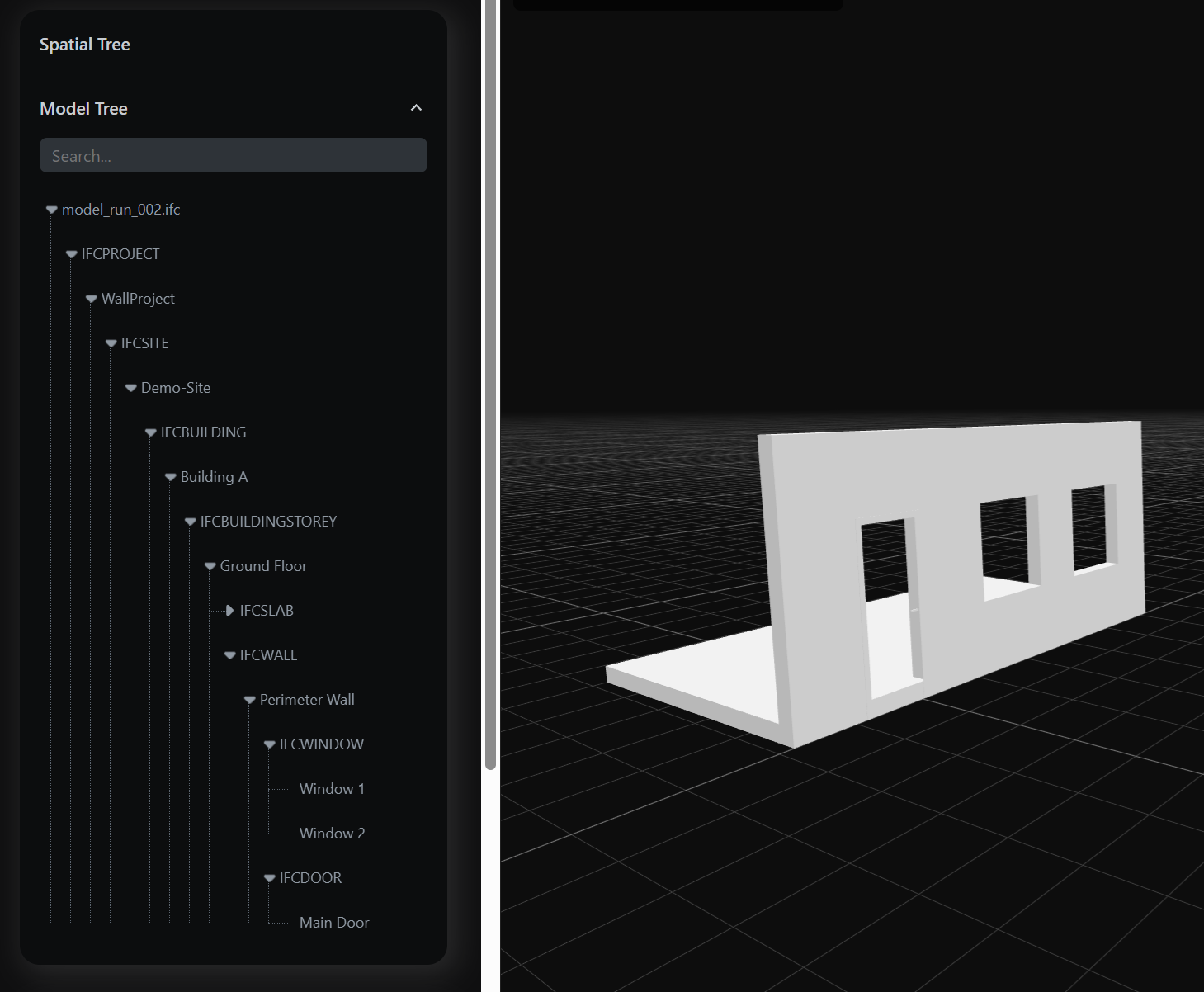}\\[3pt]
  \end{minipage}
  \hfill
  \begin{minipage}[t]{0.3\textwidth}
    \centering
    \includegraphics[width=\linewidth]{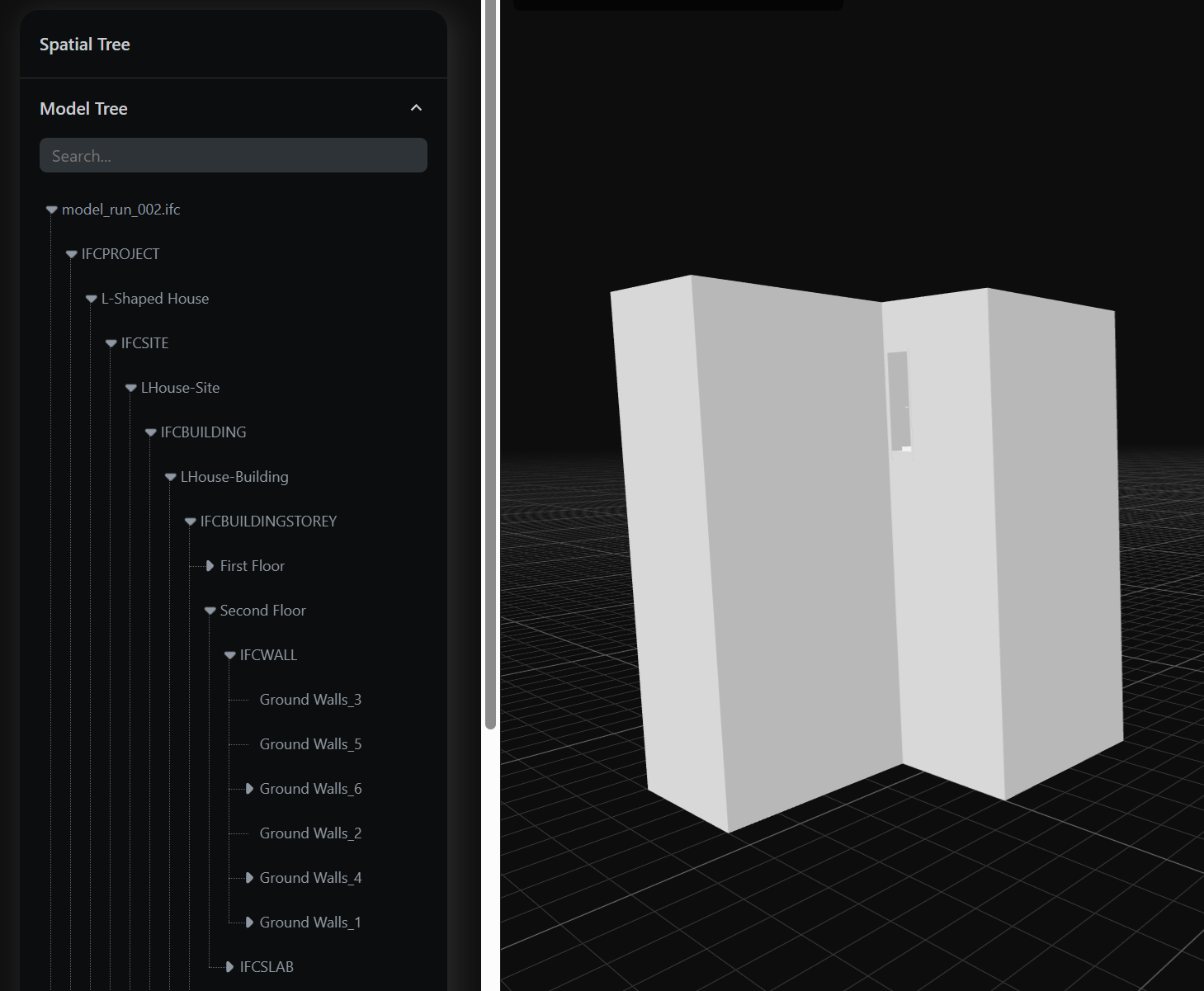}\\[3pt]
  \end{minipage}
  \hfill
  \begin{minipage}[t]{0.3\textwidth}
    \centering
    \includegraphics[width=\linewidth]{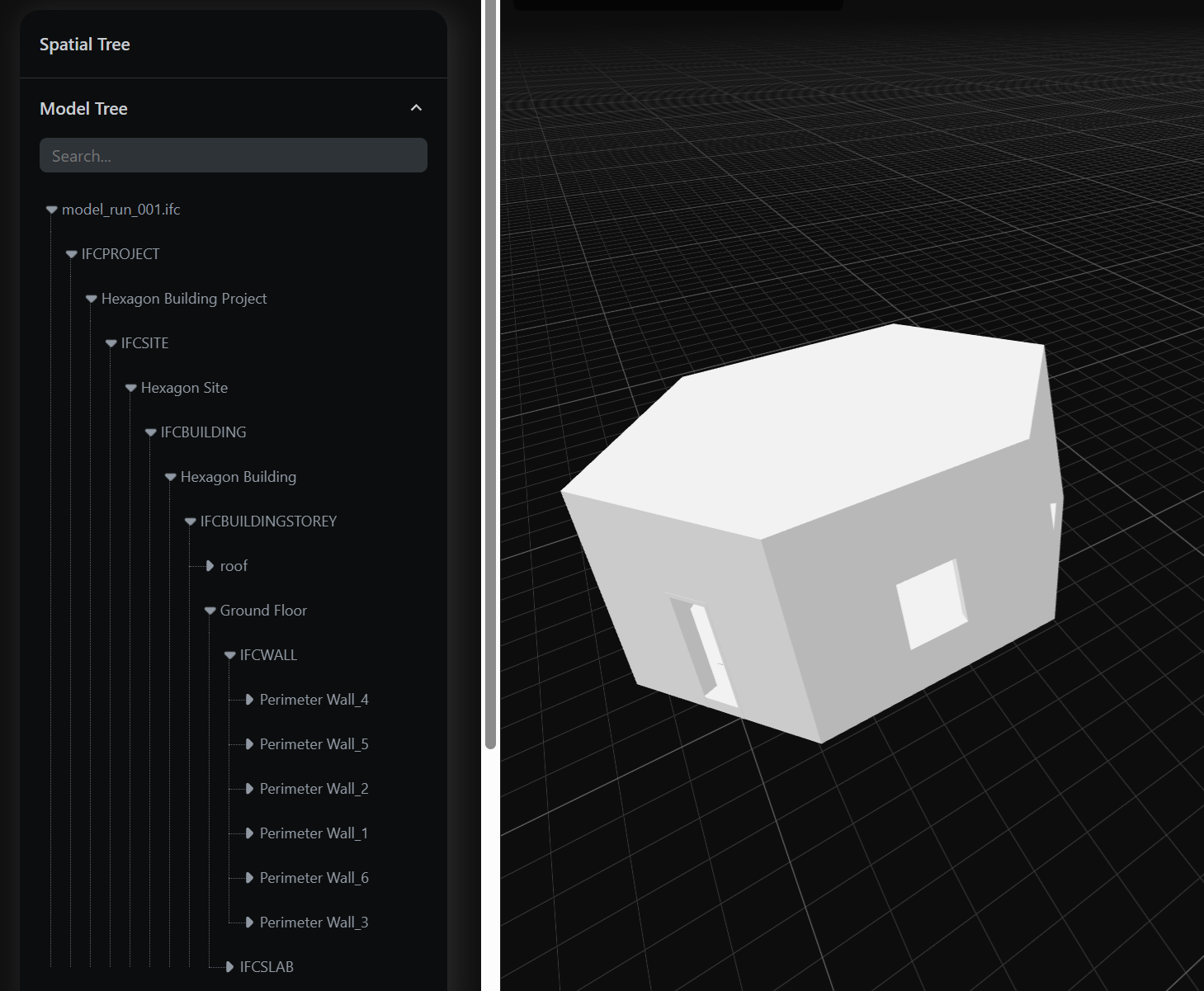}\\[3pt]
  \end{minipage}

  \caption{Representative resulting models from evaluated test cases}
  \label{fig:tc_resulting_models}
\end{figure}

The new test cases were designed to test correct tool usage, project initialization, and georeferencing capabilities. The remaining four adapted cases are selected to cover a representative range of spatial, semantic, and geometric complexity suitable for evaluating the simplified agent.
Each test case is defined as a JSON object specifying the prompt and a set of quantitative rules using IfcOpenShell’s selector syntax. The complete \texttt{test-case.json} specification is provided in the appendix for reproducibility.
The results of all 30 runs (six test cases, five repetitions each) are summarised in two tables and representative resulting models are shown in Figure~\ref{fig:tc_resulting_models}. Table~\ref{tab:agent_metrics} reports agent- and system-level metrics, including reasoning steps, token usage, number of tool calls, and tool-success rate. Table~\ref{tab:model_metrics} summarises model-level metrics, comprising schema and industry-practice validation, rule conformity, and semantic correctness ratings based on manual reviewing.

\begin{table}[h]
\centering
\begin{threeparttable}
\caption{Agent Metrics}
\label{tab:agent_metrics}
\begin{tabular}{l c c c c}
\hline
\textbf{Test Case} & \textbf{Steps} & \textbf{Tool Calls} & \textbf{Tool-Success Rate (\%)} & \textbf{Tokens Total per Run (K)} \\
\hline
tc\_du-et-al\_4 & 26.8 & 25.8 & 100.0 & 685.661  \\
tc\_du\_et\_al\_3 & 17.6 & 16.6 & 100.0 & 352.427  \\
tc\_du\_et\_al\_5 & 19.2 & 18.2 & 100.0 & 421.361  \\
tc\_du\_et\_al\_6 & 17.8 & 16.8 & 100.0 & 345.733  \\
tc\_new\_1 & 13.6 & 12.6 & 100.0 & 203.476  \\
tc\_new\_2 & 16.6 & 15.6 & 100.0 & 280.453  \\
\hline
\end{tabular}

\begin{tablenotes}
\footnotesize
Values represent the Average over 5 runs
\end{tablenotes}
\end{threeparttable}
\end{table}

\begin{table}[!hb]
\centering
\begin{threeparttable}
\caption{Model Metrics}
\label{tab:model_metrics}

\begin{tabular}{l c c c c}
\hline
\textbf{Test Case} & \textbf{Manual Review} & \textbf{Model Success (\%)} & \textbf{IFC Schema (\%)} & \textbf{Industry Practice (\%)} \\
\hline
tc\_du-et-al\_4 & 3.5 & 95.0 & 100.0 & 100.0 \\
tc\_du\_et\_al\_3 & 2.7 & 40.0 & 100.0 & 100.0 \\
tc\_du\_et\_al\_5 & 3.9 & 48.9 & 100.0 & 100.0 \\
tc\_du\_et\_al\_6 & 4.8 & 100.0 & 100.0 & 100.0 \\
tc\_new\_1 & 4.2 & 100.0 & 100.0 & 100.0 \\
tc\_new\_2 & 5.0 & 100.0 & 100.0 & 100.0 \\
\hline
\end{tabular}

\begin{tablenotes}
\footnotesize
Values represent the Average over 5 runs
\end{tablenotes}
\end{threeparttable}
\end{table}

\section{Discussion}

The analysis of existing agentic BIM workflows and MCP-based systems shows that a consistent set of core capabilities is repeatedly required across studies. As summarised in Table~\ref{tab:capabilites}, these capabilities are widely implemented but typically realised directly within the BIM authoring tool or API, resulting in strong coupling between agentic logic, tool-specific data structures, and local execution environments. Our prototype demonstrates that these capabilities can instead be separated and re-implemented as an MCP-based microservice architecture, enabling greater isolation and exchangeability.

The proposed microservice-based reference architecture constitutes a central contribution of this work. In contrast to the predominantly monolithic and tool-specific MCP implementations identified in the state of the art (Table~\ref{tab:mcp_servers}), the design establishes explicit adapter contracts and containerised execution endpoints. This architectural separation enables conceptual independence from the selected BIM authoring tool or API and reduces the tight coupling characteristic of existing approaches. The evaluation, based on a simplified re-implementation of the Text2BIM workflow, further illustrates that modelling tasks can be achieved with a comparatively minimal agent setup. A structured MCP toolset and the identified core capabilities allowed the agent to perform simple to medium-complex operations. These were achieved with limited prompting and without backend-specific logic, indicating reduced agent complexity and potential for generalisation. The limitations observed in the prototype largely reflect current LLM constraints rather than shortcomings of the architecture. Missing spatial and geometric reasoning capabilities -- such as imprecise coordinate handling or difficulties with relative placements -- remain general challenges of contemporary agentic systems and are not specific to BIM or MCP-based interaction.

The evaluation presented in this work is intentionally limited in scope. Based on a single BIM backend (IfcOpenShell) and a single-agent configuration, it neither includes comparative studies against monolithic MCP server implementations nor microservice-specific benchmarks. As a result, properties such as modularity, execution isolation, and service exchangeability are discussed primarily as architectural properties arising from the proposed decomposition and interfaces. Architecturally, constraints mainly arise from the execution environment. The BIM backend must run headlessly and expose a stable API; while such requirements are easily satisfied for open-source libraries, commercial authoring tools usually require additional remote-bridge adapters or thin client-server connectors, introducing further integration effort. These requirements remain contained within the execution service, preserving backend neutrality at the MCP layer. Finally, the sequential, stateless interaction model restricts parallel tool calls and concurrent model modifications, requiring batch operations to be encapsulated within individual tool invocations. While this simplifies reasoning and reproducibility, it may limit scalability for larger modelling tasks. Evenly complementary to widely established S3-compatible object storage, more performant, state-of-the-art storage backends may be more suitable for computing-critical tasks and are left for future investigation. Further the modular microservice decomposition introduces additional overhead and coordination effort although it simultaneously improves transparency and reusability.

\section{Conclusion and Outlook}

This work introduced a modular reference architecture for MCP servers enabling agentic BIM interaction. Based on a systematic analysis of recurring capabilities in existing agentic BIM workflows, the architecture demonstrates that retrieval, modification, and generation tasks can be implemented in a reusable, BIM-API-agnostic, and isolated manner. The prototype built around an IfcOpenShell execution backend shows that isolated, containerised execution enabled by explicit adapter contracts and endpoints can decouple LLM agents from BIM-authoring-tool-specific structures, supporting flexible and reproducible agentic workflows.

Several essential extensions remain. First, the validation of the last three core capabilities not realised in this work: \textit{Model Validation \& Quality Assurance}, \textit{Multi-Modal Inputs \& Outputs} and \textit{Information Retrieval \& Exploration} is pending. In particular, recent knowledge-graph-based approaches offer promising directions for semantic reasoning, constraint checking, and retrieval. Second, handling parallel tool calls and concurrent model modifications in the proposed stateless interaction model requires further development. Third, validating the execution-isolation concept with a non-open-source BIM authoring tool is essential to demonstrate backend-agnostic interoperability.



\section{Acknowledgements}
Claims expressed in this article do not necessarily have to coincide with the positions of the BBR/BBSR.


\clearpage
\section{References} 
\printbibliography[heading=none]
\clearpage
\section*{Appendix}
\setcounter{subsection}{0}
\renewcommand\thesubsection{\Roman{subsection}}

\subsection{MCP implementations in BIM context}

\begin{table}[h]
\centering
\begin{threeparttable}
\caption{MCP server implementations for agentic BIM interaction}
\label{tab:mcp_servers_appendix}

\footnotesize
\begin{tabular}{
  l
  l
  c
  c
  p{2cm}
  p{5.0cm}
}
\hline
\textbf{Name} &
\textbf{BIM Auth.} &
\textbf{Open-BIM} &
\textbf{Monolith} &
\textbf{Protocol} &
\textbf{Link} \\
\hline

Bonsai-MCP & Bonsai & $\checkmark$ & $\checkmark$ & stdio &
\url{https://github.com/JotaDeRodriguez/Bonsai_mcp/}\tnote{***} \\

ifcMCP & IFCOpenShell & $\checkmark$ & $\checkmark$ & HTTP (stream.) &
\url{https://github.com/smartaec/ifcMCP}\tnote{***} \\

MCP4IFC\tnote{*} & Bonsai & $\checkmark$ & $\checkmark$ & stdio &
\url{https://github.com/Show2Instruct/ifc-bonsai-mcp}\tnote{***} \\

Fragment MCP & Web-IFC & $\checkmark$ & $\checkmark$ & stdio &
\url{https://github.com/helenkwok/openbim-mcp}\tnote{***} \\

Tapir-MCP & Archicad & $\times$ & $\checkmark$ & stdio &
\url{https://github.com/SzamosiMate/tapir-archicad-MCP}\tnote{***} \\

revit-mcp & Revit & $\times$ & $\checkmark$ & stdio &
\url{https://github.com/revit-mcp/revit-mcp}\tnote{***} \\

xml.Revit.MCP & Revit & $\times$ & $\checkmark$ & stdio &
\url{https://github.com/zedmoster/revit-mcp}\tnote{***} \\

Revit MCP (Beta)\tnote{**} & Revit & $\times$ & ? & Unknown &
\url{https://www.autodesk.com/solutions/autodesk-ai/autodesk-mcp-servers}\tnote{***} \\
\hline
\end{tabular}

\begin{tablenotes}
\footnotesize
\item[*] \parencite{nithyanantham_2025_MCP4IFCIFCBasedBuilding}
\item[**] Only beta announced.
\item[***] Visited 25.11.2025
\end{tablenotes}

\end{threeparttable}
\end{table}

\subsection{ Model - Agent interaction Loop }
\begin{figure}[h]
  \centering
  \includegraphics[width=\figwidth]{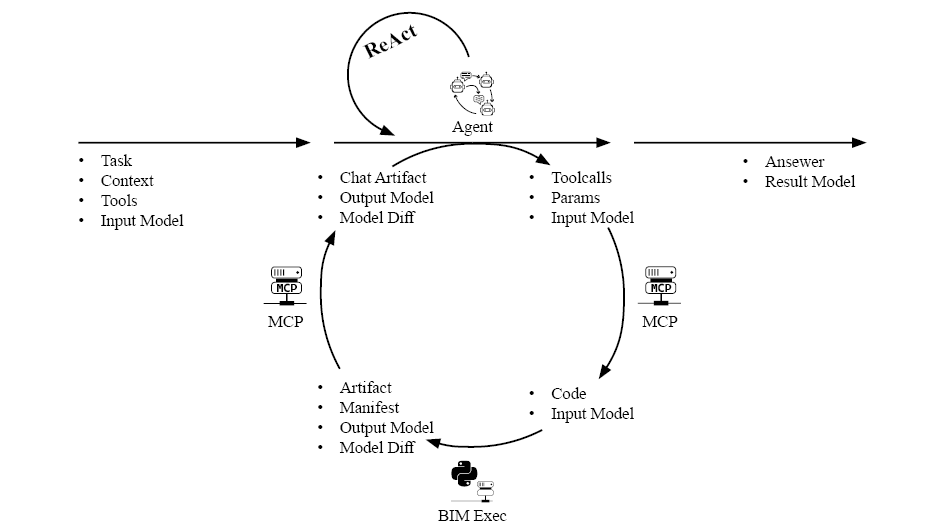}
  \caption{Conceptual interaction flow between the agent, MCP server, and BIM Execution Service.}
  \label{fig:modelexecflow}
\end{figure}

Figure~\ref{fig:modelexecflow} illustrates the interaction loop between the agent, the MCP server, and the BIM Execution Service.

\subsection{Test Cases}

\begin{table}[h]
\setlength{\tabcolsep}{12pt} 
\centering
\caption{Manual review scale for LLM-generated BIM models.}
\label{tab:manual_review_scale}
\begin{tabular}{l l l}
\hline
\textbf{Level} & \textbf{Rating} & \textbf{Description} \\
\hline
Level 1 & {\dejavusans ★☆☆☆☆} & No valid model; IFC structurally invalid or unreadable. \\
Level 2 & {\dejavusans ★★☆☆☆} & \textit{Level 1} + Syntactically valid IFC with basic spatial hierarchy. \\
Level 3 & {\dejavusans ★★★☆☆} & \textit{Level 2} + Correct element usage and topology (containment, openings). \\
Level 4 & {\dejavusans ★★★★☆} & \textit{Level 3} + Spatial correctness: plausible placement of doors, windows, storeys. \\
Level 5 & {\dejavusans ★★★★★} & \textit{Level 4} + Geometric correctness and consistent semantics. \\

\end{tabular}
\end{table}

\subsubsection*{Test Case: New\_1}
\begin{figure}[!htbp]

  \centering

  {\small\itshape
    JSON specification (top) and image of the resulting model after run 002 (bottom).
  \par}
  \vspace{0.6em}

  \scalebox{0.7}{%
  \begin{minipage}{\linewidth}
    \centering
\begin{minted}{json}
{
  "tc_new_1": {
    "prompt": "Create a wall with a length of 7 meters and a height of 3 meters; and 2 windows and 1 door to the wall.",
    "success_criteria": {
      "element_existence": {
        "IfcWall": 1,
        "IfcWindow": {
          "min": 2,
          "max": 2
        },
        "IfcDoor": 1
      },
      "element_features": {
        "wall_dimensions": {
          "selector": "IfcWall, Qto_WallBaseQuantities.Length=7, Qto_WallBaseQuantities.Height=3"
        }
      }
    }
  }
}
\end{minted}
  \end{minipage}
  }

  \vspace{0.8em}
  {\color{gray}\rule{\figwidth}{0.4pt}}
  \vspace{0.8em}

  \begin{minipage}{\figwidth}
    \centering
    \includegraphics[width=\figwidth]{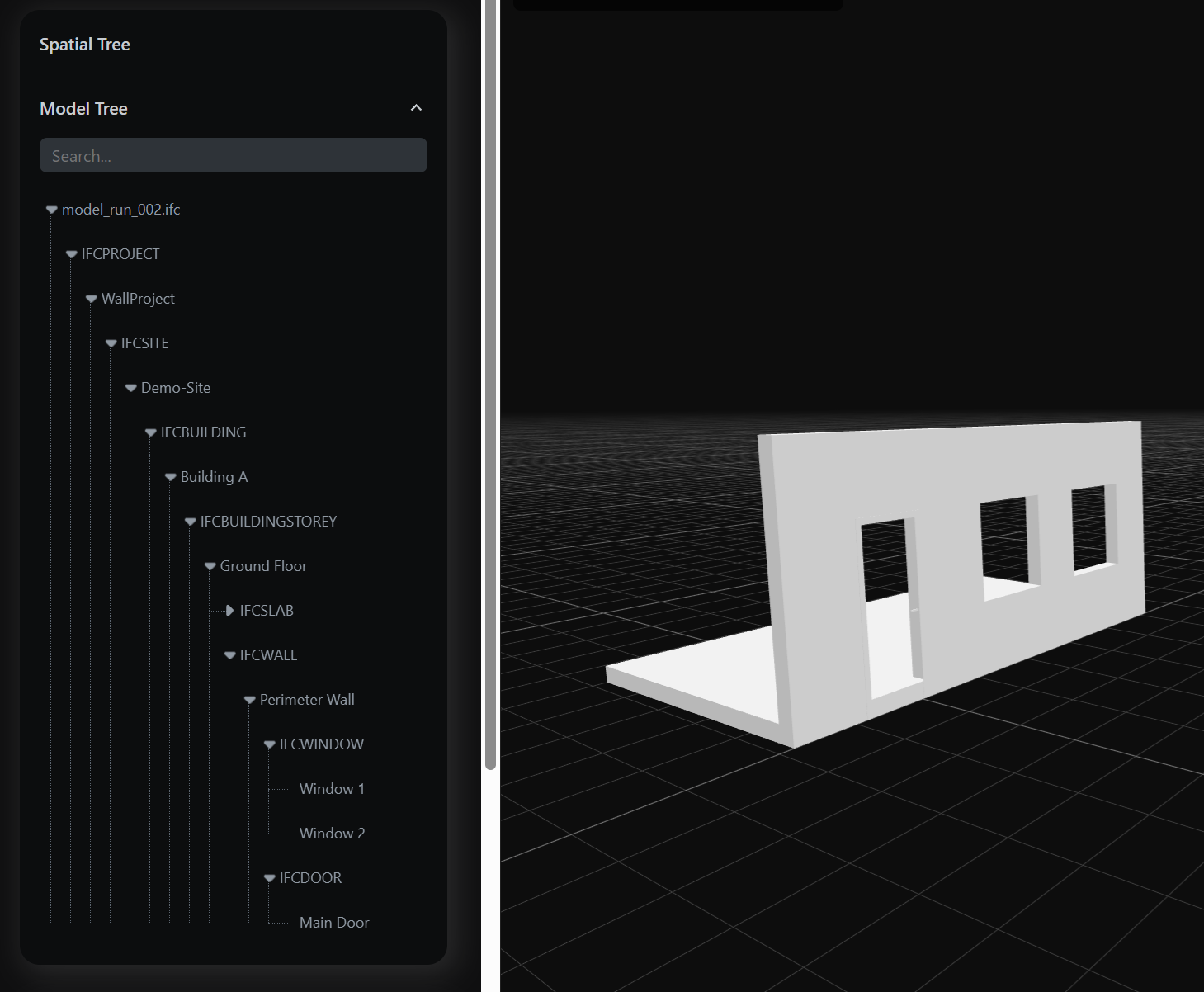}
  \end{minipage}

\end{figure}

\begin{figure}[ht]
  \subsubsection*{Test Case: New\_2}
  \centering

  {\small\itshape
    JSON specification (top) and image of the resulting model after run 001 (bottom).
  \par}
  \vspace{0.6em}

  \scalebox{0.7}{%
  \begin{minipage}{\linewidth}
    \centering
\begin{minted}{json}
{
  "tc_new_2": {
    "prompt": "Create a simple IFC project for a new Alan Turing Institute on the Turing Way 1 in Cambridge, UK. Add a Site called 'Turing Place'. Create a Building for the Institute named 'Alan Turing Institute for AI' with one storey ('Ground Floor') and one room; The dimensions of the building are 20m x 30m; add perimeter walls, slabs, and a roof. Add at least 3 windows, and add one door. Add the in the first wall at one-third of its length. Give a link to view the result.",
    "success_criteria": {
      "element_existence": {
        "IfcBuildingStorey": {
          "min": 2
        },
        "IfcBuilding": 1,
        "IfcDoor": {
          "min": 1,
          "max": null
        },
        "IfcWindow": {
          "min": 3,
          "max": null
        },
        "IfcRoof": 1,
        "IfcSlab": {
          "min": 1,
          "max": null
        },
        "IfcSpace": {
          "min": 1,
          "max": null
        },
        "IfcWall": {
          "min": 4,
          "max": null
        }
      },
      "element_features": {
        "site_named": "IfcSite, Name=\"Turing Place\"",
        "building named": "IfcBuilding, Name=\"Alan Turing Institute for AI\"",
        "storey_named": "IfcBuildingStorey, Name=\"Ground Floor\"",
        "wall_dimensions_long": {
          "selector": "IfcWall, Qto_WallBaseQuantities.Length=20",
          "min": 2
        },
        "wall_dimensions_short": {
          "selector": "IfcWall, Qto_WallBaseQuantities.Length=30",
          "min": 2
        },
        "storey_height": {
          "selector": "IfcBuildingStorey, Name=\"roof\", Elevation<\"2.5\"",
          "min": 0
        },
        "building footprint between 540 and 660": {
          "selector": "IfcBuilding, Qto_BuildingBaseQuantities.FootprintArea>=540, Qto_BuildingBaseQuantities.FootprintArea<=660",
          "min": 1
        }
      }
    }
  }
}
\end{minted}
  \end{minipage}
  }

  \vspace{0.8em}
  {\color{gray}\rule{\figwidth}{0.4pt}}
  \vspace{0.8em}

  \begin{minipage}{\figwidth}
    \centering
    \includegraphics[width=\figwidth]{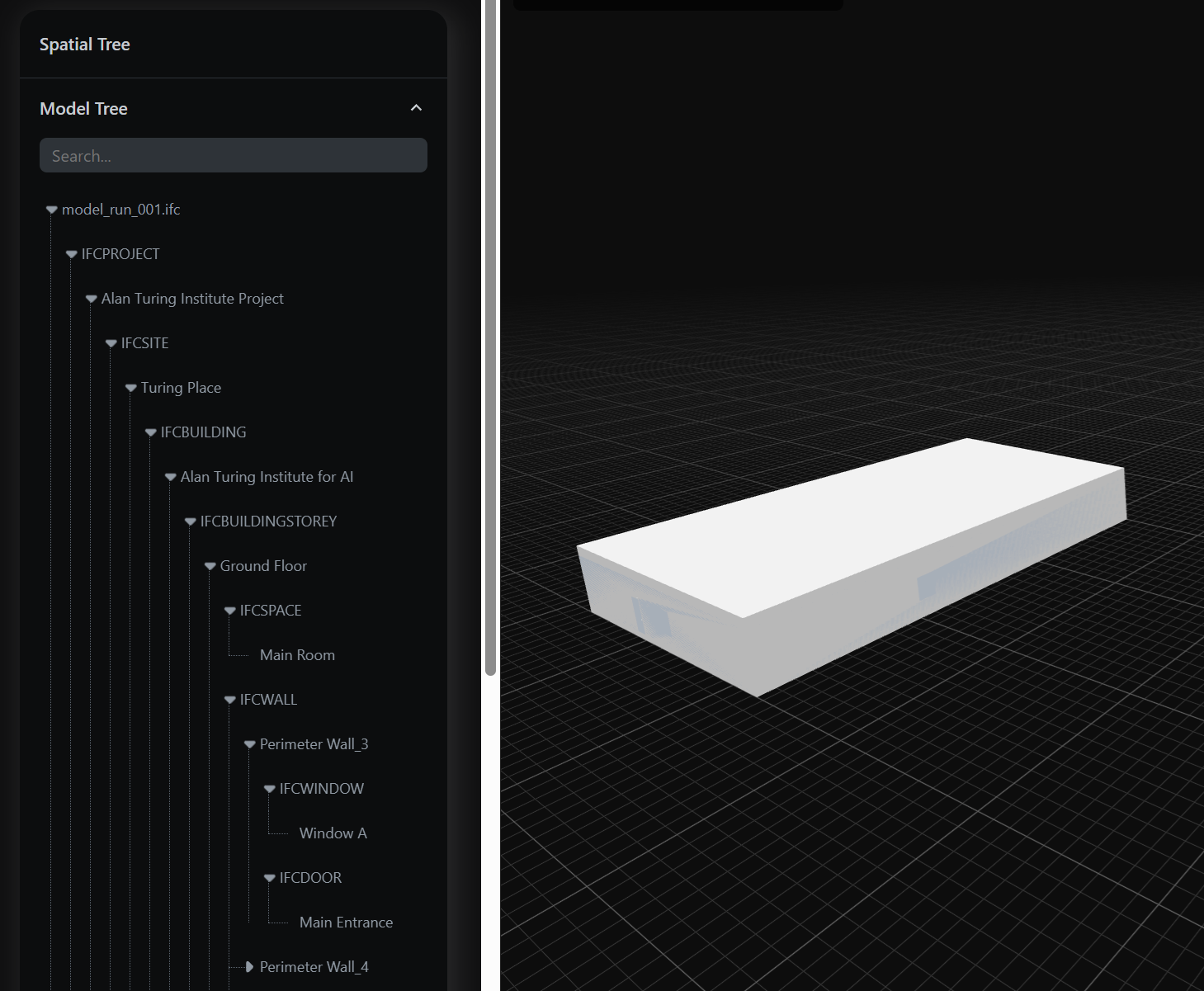}
  \end{minipage}

\end{figure}

\begin{figure}[ht]
  \subsubsection*{Test Case: du\_et\_al\_3}
  \centering

  {\small\itshape
    JSON specification (top) and image of the resulting model after run 002 (bottom).
  \par}
  \vspace{0.6em}

  \scalebox{0.7}{%
  \begin{minipage}{\linewidth}
    \centering
\begin{minted}{json}
{
  "tc_du_et_al_3": {
    "prompt": "Create a basic 3D model of a four-story residential house with overall footprint dimensions 5m by 3m.",
    "success_criteria": {
      "element_existence": {
        "IfcBuilding": 1,
        "IfcBuildingStorey": {
          "min": 5,
          "max": 5
        },
        "IfcWall": {
          "min": 16,
          "max": null
        },
        "IfcSlab": {
          "min": 5,
          "max": null
        },
        "IfcRoof": {
          "min": 1,
          "max": null
        },
        "IfcDoor": {
          "min": 1,
          "max": null
        },
        "IfcWindow": {
          "min": 1,
          "max": null
        }
      },
      "element_features": {
        "walls_length_5m": {
          "selector": "IfcWall, Qto_WallBaseQuantities.Length=5",
          "min": 8,
          "max": null
        },
        "walls_length_3m": {
          "selector": "IfcWall, Qto_WallBaseQuantities.Length=3",
          "min": 8,
          "max": null
        },
        "storey_perimeter_5mx3m": {
          "selector": "IfcSlab, Qto_SlabBaseQuantities.Perimeter=16",
          "min": 4
        }
      }
    }
  }
}
\end{minted}
  \end{minipage}
  }

  \vspace{0.8em}
  {\color{gray}\rule{\figwidth}{0.4pt}}
  \vspace{0.8em}

  \begin{minipage}{\figwidth}
    \centering
    \includegraphics[width=\figwidth]{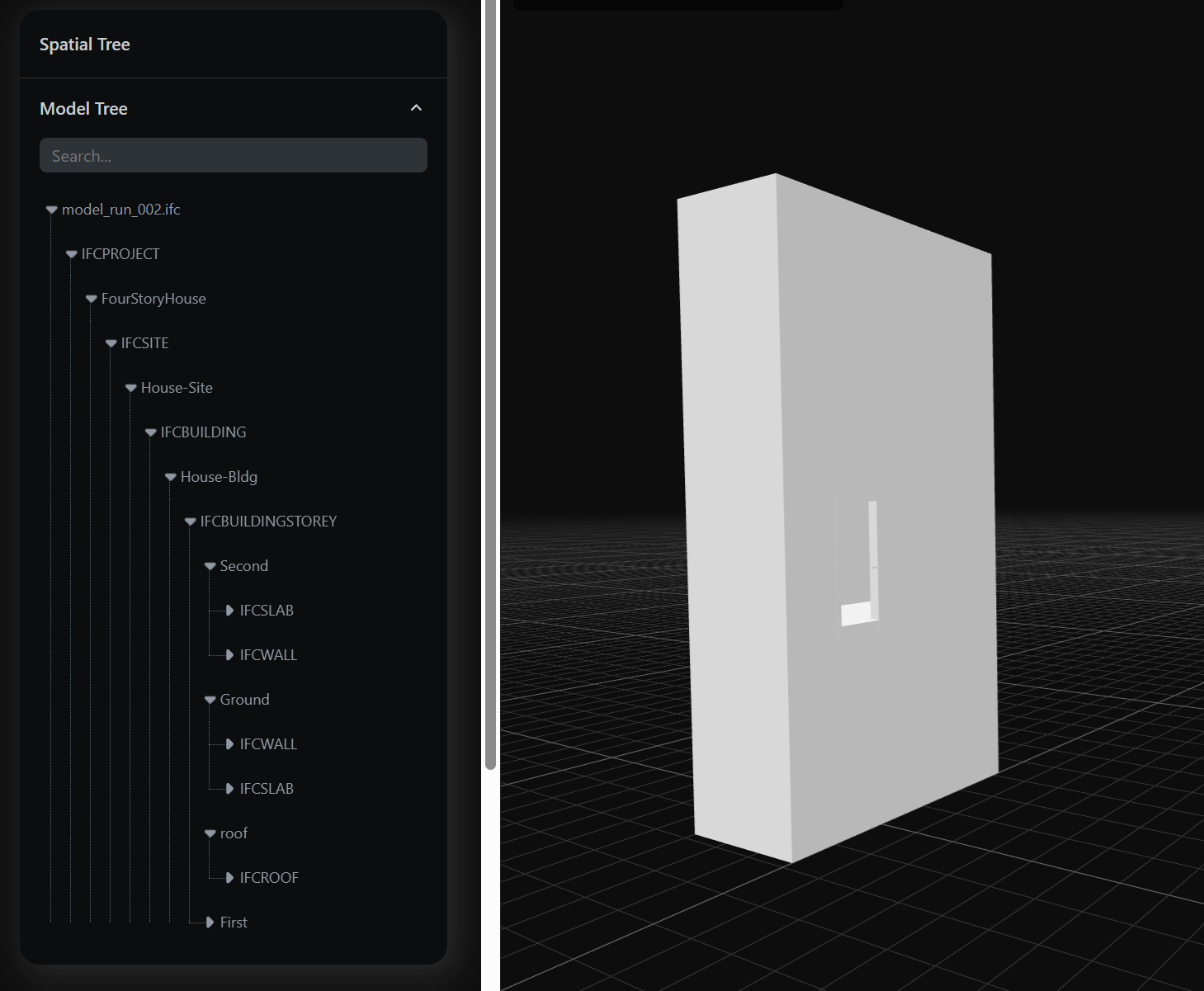}
  \end{minipage}

\end{figure}

\begin{figure}[ht]
  \subsubsection*{Test Case: du\_et\_al\_4}
  \centering

  {\small\itshape
    JSON specification (top) and image of the resulting model after run 004 (bottom).
  \par}
  \vspace{0.6em}

  \scalebox{0.7}{%
  \begin{minipage}{\linewidth}
    \centering
\begin{minted}{json}
{
  "tc_du-et-al_4": {
    "prompt": "Create a single-story residential house with a total floor area of 120 square meters. Include three bedrooms, two bathrooms, a kitchen, and a living room. Label rooms <function>_n lower case e.g. bedroom_1. The house should have a flat roof, and incorporate at least four windows and one main entrance door. Lable perimeter walls as perimeter_wall_n",
    "success_criteria": {
      "element_existence": {
        "IfcBuilding": 1,
        "IfcBuildingStorey": 2,
        "IfcSpace": {
          "min": 7,
          "max": null
        },
        "IfcWindow": {
          "min": 4,
          "max": null
        },
        "IfcDoor": {
          "min": 1,
          "max": null
        },
        "IfcRoof": {
          "min": 1,
          "max": null
        },
        "IfcSlab": {
          "min": 1,
          "max": null
        }
      },
      "element_features": {
        "3_bedrooms_exist": {
          "selector": "IfcSpace, Name=/bed*/",
          "min": 3
        },
        "2_bathrooms_exist": {
          "selector": "IfcSpace, Name=/bath*/",
          "min": 2
        },
        "kitchen_exists": {
          "selector": "IfcSpace, Name=/kitchen*/"
        },
        "living_room_exists": {
          "selector": "IfcSpace, Name=/living*/"
        },
        "floor_area_~120m²": {
          "selector": "IfcBuildingStorey, Qto_BuildingStoreyBaseQuantities.GrossFloorArea=120"
        }
      }
    }
  }
}
\end{minted}
  \end{minipage}
  }

  \vspace{0.8em}
  {\color{gray}\rule{\figwidth}{0.4pt}}
  \vspace{0.8em}

  \begin{minipage}{\figwidth}
    \centering
    \includegraphics[width=\figwidth]{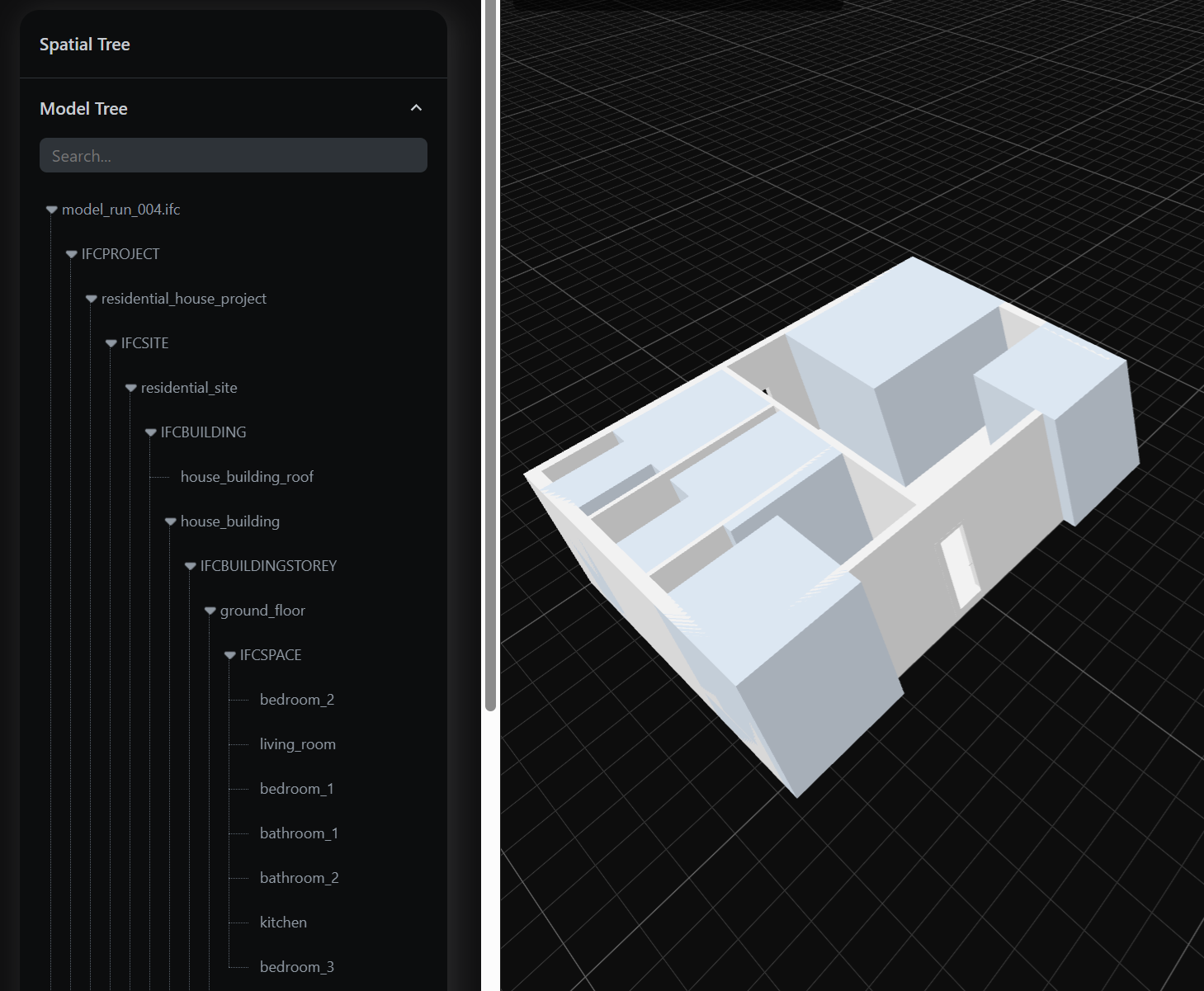}
  \end{minipage}

\end{figure}

\begin{figure}[ht]
  \subsubsection*{Test Case: du\_et\_al\_5}
  \centering

  {\small\itshape
    JSON specification (top) and image of the resulting model after run 002 (bottom).
  \par}
  \vspace{0.6em}

  \scalebox{0.7}{%
  \begin{minipage}{\linewidth}
    \centering
\begin{minted}{json}
{
  "tc_du_et_al_5": {
    "prompt": "Create an 3-story L-shaped house with each leg of the L being 8 meters long and 4 meters wide. Place a door at the corner of the L and a window on each side of the L.",
    "success_criteria": {
      "element_existence": {
        "IfcBuilding": 1,
        "IfcBuildingStorey": {
          "min": 4,
          "max": 4
        },
        "IfcWall": {
          "min": 18,
          "max": null
        },
        "IfcDoor": {
          "min": 1,
          "max": 1
        },
        "IfcWindow": {
          "min": 2,
          "max": 2
        },
        "IfcSlab": {
          "min": 3,
          "max": null
        },
        "IfcRoof": {
          "min": 1,
          "max": null
        }
      },
      "element_features": {
        "some_walls_length_8m": {
          "selector": "IfcWall, Qto_WallBaseQuantities.Length=8",
          "min": 6,
          "max": null
        },
        "some_walls_length_4m": {
          "selector": "IfcWall, Qto_WallBaseQuantities.Length=4",
          "min": 6,
          "max": null
        }
      }
    }
  }
}
\end{minted}
  \end{minipage}
  }

  \vspace{0.8em}
  {\color{gray}\rule{\figwidth}{0.4pt}}
  \vspace{0.8em}

  \begin{minipage}{\figwidth}
    \centering
    \includegraphics[width=\figwidth]{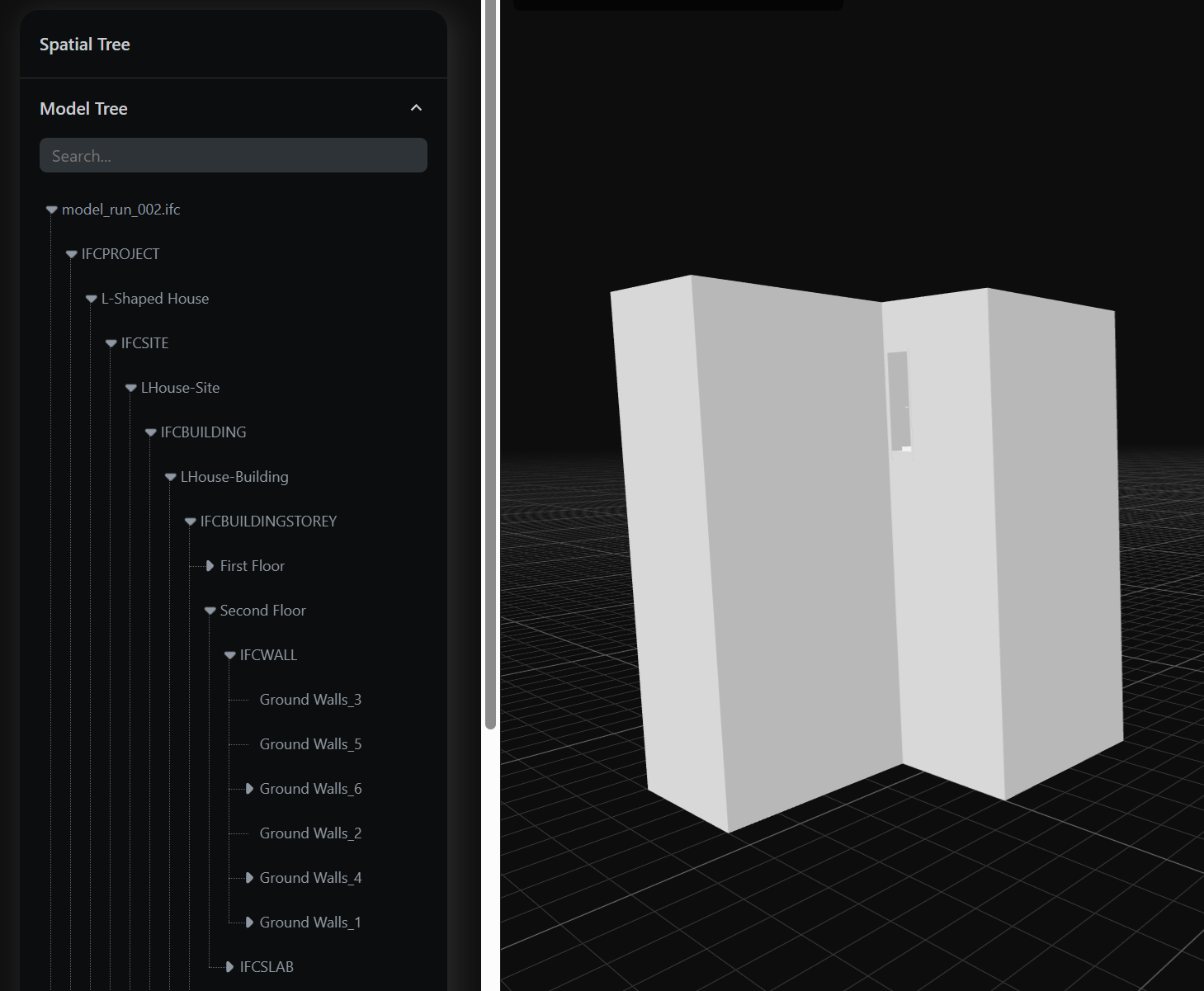}
  \end{minipage}

\end{figure}

\begin{figure}[ht]
  \subsubsection*{Test Case: du\_et\_al\_6}
  \centering

  {\small\itshape
    JSON specification (top) and image of the resulting model after run 001 (bottom).
  \par}
  \vspace{0.6em}

  \scalebox{0.7}{%
  \begin{minipage}{\linewidth}
    \centering
\begin{minted}{json}
{
  "tc_du_et_al_6": {
    "prompt": "Design a building with a hexagonal footprint. Each side of the hexagon should be 5 meters. Add a slab for the floor and a flat roof. Include a door on one side and a window on each of the other sides.",
    "success_criteria": {
      "element_existence": {
        "IfcBuilding": 1,
        "IfcBuildingStorey": {
          "min": 2,
          "max": 2
        },
        "IfcWall": {
          "min": 6,
          "max": null
        },
        "IfcSlab": {
          "min": 1,
          "max": null
        },
        "IfcRoof": {
          "min": 1,
          "max": null
        },
        "IfcDoor": 1,
        "IfcWindow": 5
      },
      "element_features": {
        "hex_walls_length_5m": {
          "selector": "IfcWall, Qto_WallBaseQuantities.Length>=\"4.8\", Qto_WallBaseQuantities.Length<=\"5.2\"",
          "min": 6,
          "max": null
        },
        "hex_footprint_by_perimeter": {
          "selector": "IfcSlab, Qto_SlabBaseQuantities.GrossPerimeter>=\"29.8\", Qto_SlabBaseQuantities.GrossPerimeter<=\"30.2\""
        },
        "hex_footprint_by_area": {
          "selector": "IfcSlab, Qto_SlabBaseQuantities.GrossArea>=60, Qto_SlabBaseQuantities.GrossArea<=70"
        }
      }
    }
  }
}
\end{minted}
  \end{minipage}
  }

  \vspace{0.8em}
  {\color{gray}\rule{\figwidth}{0.4pt}}
  \vspace{0.8em}

  \begin{minipage}{\figwidth}
    \centering
    \includegraphics[width=\figwidth]{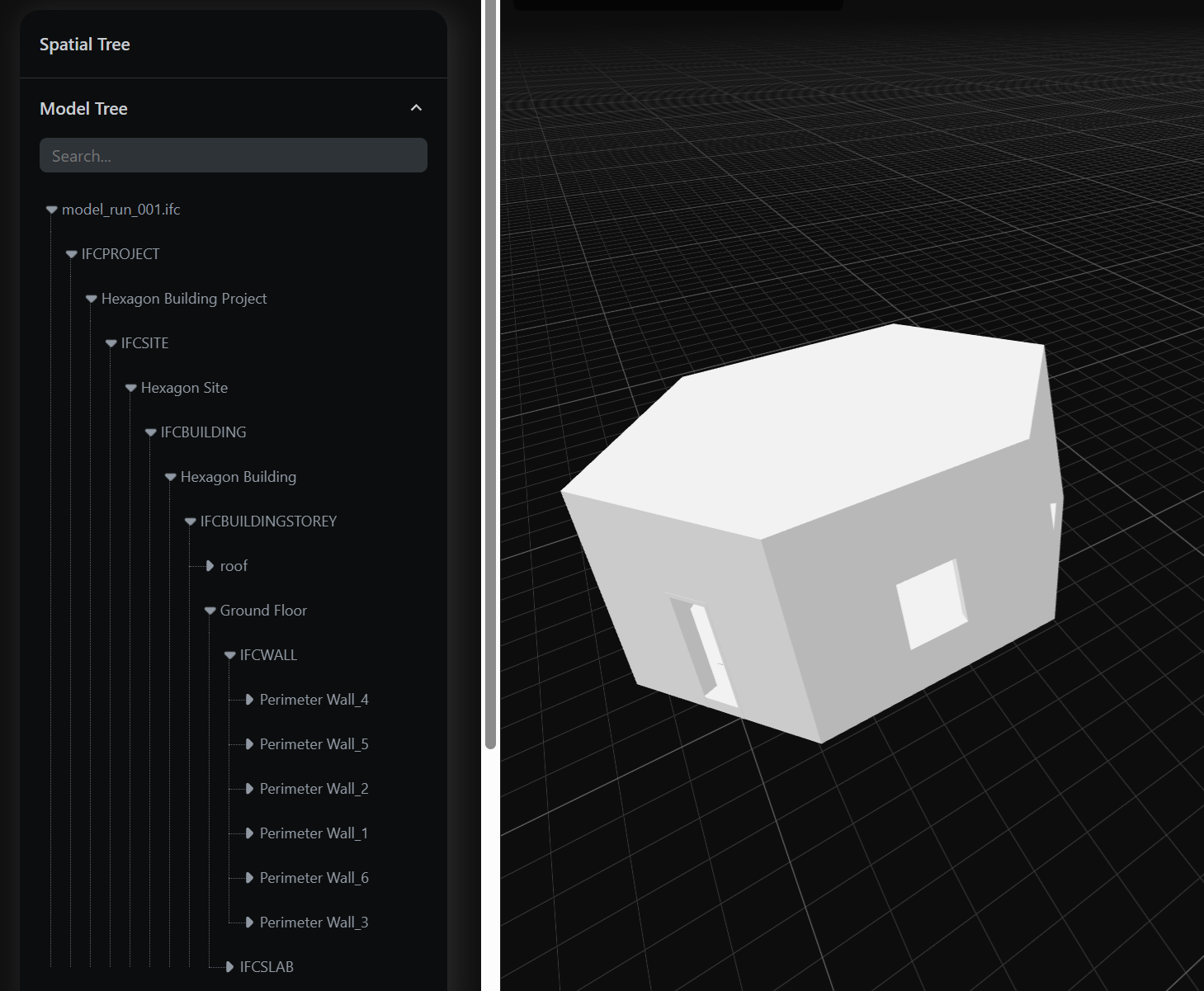}
  \end{minipage}

\end{figure}

\end{document}